\documentstyle[a4,12pt]{article}
\begin{document}

\title
{Granular Flow, Collisional Cooling and Charged Grains}

\author
{D. E. Wolf\thanks{d.wolf@uni-duisburg.de}, T. Scheffler and
  J. Sch\"afer\thanks{Present address: 
Procter \& Gamble European Services GmbH, D-65823 Schwalbach am
Taunus, Germany }\\ 
Theoretische Physik, Gerhard-Mercator-Universit\"at Duisburg\\ D-47048
Duisburg, Germany}

\maketitle

\section*{Abstract}
The non-Newtonian character of granular flow in a vertical pipe is
analyzed. The time evolution of the flow velocity and the velocity
fluctuations (or granular temperature) is derived. The steady state
velocity has a power law dependence on the pipe width with an exponent
between $3/4$ and $3/2$. The flow becomes faster the more efficient
the collisional cooling is, provided the density remains low enough.
The dependence of collisional cooling on the solid fraction, the
restitution coefficient and a possible electric charge of all grains
is discussed in detail. 

\section{Introduction}

Granular materials like dry sand are classical many particle systems
which differ significantly from solids and liquids in their dynamical
behavior. They seem to have some properties in common
with solids, like the ability of densely packed grains to sustain
shear. Other properties, like the ability to flow through a
hopper, remind of a fluid. Upon looking more closely, however, it
turns out that such similarities are only superficial: Force
localization or arching in granular packings and their plastic yield
behaviour are distictly different from solids, cluster instabilities
and nonlinear internal friction make granular flow very different from
that of ordinary, Newtonian fluids. These differences manifest
themselves often in extraordinarily strong fluctuations, which
may cause accidents, when ignored in technological applications.
This is why understanding the statistical physics of granular
materials is important and has been very fruitful (see
e.g. \cite{reviews,Wolf}). In this paper 
a few selected examples will be presented, which on the one hand
document the progress in understanding brought about by applying 
concepts from statistical physics, and on the other hand point out
some areas where important and difficult questions invite future
research.

One of the simplest geometries displaying the non-Newtonian character
of granular flow is an evacuated vertical tube through which the
grains fall. The experimental investigation is difficult, however,
because the flow depends sensitively on electric charging and
humidity \cite{Raafat}. Nevertheless the ideal uncharged dry granular
medium falling 
in vacuum through a vertical pipe is an important reference case and a
natural starting point for computer simulations. 

Using this idealized example we shall show that the properties of 
granular flow can be explained, if two essential physical ingredients
are understood: The interaction of the granular flow with the container
walls and the phenomenon of collisional cooling. This technical term
draws an analogy between the disordered relative motion of the
agitated grains and the thermal motion of gas molecules. In a
granular ``gas'', differently from a molecular gas, the relative
motion of the grains is reduced in every collision due to the
irreversible loss of kinetic energy to the internal degrees of freedom
of the grains. This is called collisional cooling.

Agitated dry grains are usually electrically charged due to contact
electrification. Its effect on the dynamical behavior of granular
materials has hardly been studied so far. The reason is certainly not
lack of interest, as intentional charging is the basis of several
modern applications of granular materials in industrial processes.
The reason is that well controlled experiments with electrically
charged grains are difficult, as is the theory, because of the long
range nature of the Coulomb interaction.

One such application is the electrostatic separation of scrap plastics
into the raw materials for recycling. A similar technique is used to
separate Potassium salts, a raw material for fertilizers, from rock
salt. In a ``conditioning process'' chemically different grains get
charged oppositely. Then they fall through a condenser tower, where
they are deflected in opposite directions and hence separated.
Such a dry separation has the advantage of avoiding the environmental
damage, the old fashioned chemical separation method would cause.
Another application is powder varnishing. In order to avoid the
harmful fumes of ordinary paints, the dry pigment powder is charged
monopolarly and attracted to the grounded piece of metal to be
varnished. Once covered with the powder, the metal is heated so that
the powder melts and forms a continuous film. 

Recently a rather complete understanding has been reached, how
monopolar charging affects collisional cooling \cite{Scheffler1,Scheffler2}.
However, little is known about the influence of the charges on the
grain-wall interaction statistics.
The monopolar case is much simpler than the bipolar one:
If all grains repell each other, collisional cooling cannot lead
to the clustering instability observed for neutral grains
\cite{Goldhirsch,McNamara}. The case, 
where the grains carry charges of either sign, is much more
difficult, because the clustering instability might even be enhanced.
It has not been investigated yet.

\section{Why the laws of Hagen, Poisseuille and Bagnold fail
  for granular pipe flow}

Since the flow through a vertical pipe is such a basic example, it has
been addressed many times, including some classical work from the last
century. Here we remind the reader of some of the most elementary
ideas and results concerning pipe flow, and at the same time show,
where they fail. The general situation is much more complex, as we are
going to point out in the subsequent sections.

Force balance requires
that the divergence of the stress tensor $\sigma_{ij}$ compensates the
weight per unit volume in the steady state of the flowing material:
\begin{equation}
\partial_x \sigma_{zx} + \partial_z \sigma_{zz} = - mgn ,
\label{force_balance}
\end{equation}
where $m$ denotes the molecular or grain mass, $n$ the number density
of molecules or grains and $g$ the gravitational acceleration. The
partial derivative in the vertical (the $z$-) direction vanishes because
of translational invariance along the pipe. $\partial_x$ denotes the
partial derivative in the transversal direction.\footnote{For the sake
  of transparency the equations are given for the two dimensional case
  in this section.}

In Newtonian or simple liquids the stress tensor is assumed to be
proportional to the shear rate 
\begin{equation}
\sigma_{zx} = \eta \partial_x v_z.
\label{newton}
\end{equation}
The proportionality constant $\eta$ is the viscosity.
Inserting this into (\ref{force_balance}) immediately gives the
parabolic velocity profile of Hagen-Poisseuille flow, $v_z =
v_{\rm max} - (mgn/2\eta) x^2$. No-slip boundary conditions then
imply that the flow velocity averaged over the cross section of the
pipe scales like ${\bar v}\propto W^2$.

According to kinetic theory the viscosity $\eta$ is proportional
to the thermal velocity. In lowest order the thermal motion of liquid
molecules is independent of the average
flow velocity. It is given by the coupling of the liquid to a heat bath.
For a granular gas, the thermal 
velocity must be replaced by a typical relative velocity ${\delta v}$ of the 
grains. Due to collisional cooling
${\delta v}$ would drop to zero, if there was no flow.
This is the most important difference between liquid and granular flow.
It shows, that for a granular gas the collision rate between 
the grains, and hence the viscosity cannot be regarded
as independent of the average flow velocity in lowest order.
Bagnold \cite{Bagnold}
argued that the typical relative motion should be proportional
to the absolute value of the shear rate, $\eta \propto \delta v
\propto |\partial_x v_z|$,  so that
\begin{equation}
\sigma_{zx} \propto |\partial_x v_z| \partial_x v_z.
\end{equation}
Inserting this into (\ref{force_balance}) leads to the result, that
the average flow velocity must scale with the pipe diameter as
${\bar v}\propto W^{3/2}$.

However, Bagnold's argument ignores, that there is a second
characteristic velocity in the system, which is $\sqrt{gd}$,
where $d$ is the diameter of the grains. 
It enters due to the nonlinear coupling between the flow
velocity and the irregular grain motion, as we are going to point out
in the next section. Hence, for granular flow through a vertical pipe,
the viscosity is 
a function of both the average flow velocity and $\sqrt{gd}$. This
will change the scaling of ${\bar v}$ with the diameter of the pipe, of course.

Very little is known about the flow of dry granular materials at high
solid fractions, where the picture of gas-like dynamics, which we
employed so far, no longer applies. Hagen studied the discharge 
from a silo \cite{Hagen} and postulated, that the flow rate is not
limited by plastic deformations inside the packing but by arching at the
outlet. He assumes that the only dimensionful relevant parameters for 
outlets much larger than the grain size are $g$ and the width $W$ of the
outlet, for which we use the same notation as for the pipe diameter. 
Therefore, he concludes, that up to dimensionless prefactors
\begin{equation}
{\bar v}\propto \sqrt{gW}.
\label{Eq:Hagen}
\end{equation}
He confirmed this experimentally for the silo geometry, where
the outlet is smaller than the diameter of the container.
It is tempting to expect that this holds also for pipe flow at 
high solid fractions. However, in our computer simulations we
never observed such a behaviour, although we studied volume
fractions, which were so high, that the addition of a single 
particle would block the pipe completely. Without
investigating dense granular flow any further in
this paper, we just want to point out that Hagen's dimensional 
argument seems less plausible for a pipe than for a silo, because
important arching now occurs at any place simultaneously along the pipe, and
the dynamics of decompaction waves \cite{Luding} and plastic deformations far 
from the lower end of the pipe may well depend on the dimensionless
ratio $W/d$, for instance. This spoils the argument leading to
(\ref{Eq:Hagen}), of course.

\section{General equations of momentum and energy balance}

A vertical pipe can be viewed essentially as a one-dimensional
system, if one averages all dynamical quantities over the cross section. 
In the following we derive the time evolution of such 
cross sectional averages of the velocity and velocity fluctuations,
assuming they are constant along the pipe. This assumption ignores
the spontaneous formation of density waves, which is legitimate if the
pipe is sufficiently short. Then a homogeneous state can be maintained.
It needs not be stationary, though, and the equations we shall derive 
describe its temporal evolution.
The physical significance of this study is based on the
assumption that short sections of a long pipe are
locally homogeneous and close to the corresponding steady state. 

The translational invariance along the pipe implies that the average
velocity only has an axial component. 
Its time evolution is given by the competition of
a gain term, which is the gravitational acceleration $g$, and a loss
term due to the momentum transfer to the pipe wall. 
Here we focus on the behavior at low enough densities,
where the dynamics are dominated by collisions rather than frictional
contacts. Then the momentum transfer to the pipe wall is
proportional to the number of grain-wall collisions, $\dot N_{\rm w}$. In
each such collision the axial velocity of the colliding particle
changes by an average value $\Delta {\bar v}$. All grains are assumed to be
equal for simplicity.  Hence the average axial
velocity changes by $\Delta {\bar v}/N$ in a wall collision. The momentum
balance then reads:
\begin{equation}
\dot {\bar v} = g - \dot N_{\rm w} \Delta {\bar v}/N .
\label{E_v}
\end{equation}

More subtle is the energy balance which gives rise to an equation for
the root mean square fluctuation of the velocity,
${\delta v} = \sqrt{\langle \vec{v}^2 \rangle - \langle \vec{v} \rangle
  ^2}$. This can be regarded as the typical absolute value of relative velocities.

The rates of energy dissipation $\dot E_{\rm diss}$ 
and of change of kinetic and potential energy, $\dot E_{\rm kin}$
and $\dot E_{\rm pot}$ must add up to zero due to energy conservation,
\begin{equation}
0=\dot E_{\rm diss} + \dot E_{\rm kin} + \dot E_{\rm pot}.
\label{E_energy_balance}
\end{equation}
The change in kinetic energy per unit time is
\begin{equation}
\dot E_{\rm kin} = N m ({\bar v} \dot {\bar v}  + {\delta v} \dot {\delta v} ) , 
\label{E_E_kin}
\end{equation}
where $N$ is the total number of particles in the pipe and $m$ their
mass. The potential energy (in the absence of Coulomb
interactions between the grains) changes at a rate
\begin{equation}
\dot E_{\rm pot} = - N m g {\bar v} .
\label{E_E_pot}
\end{equation}

If only the irreversible nature of binary grain collisions is
taken into account the energy dissipation rate is proportional to the number of
binary collisions per unit time, $\dot N_{\rm g}$, times the loss of
kinetic energy in the relative motion of the collision partners,
\begin{equation}
\dot E_{\rm diss} = \dot N_{\rm g} \Delta E,
\label{E_E_diss}
\end{equation} 
with
\begin{equation}
\Delta E = \Delta (m {\delta v}^2/2) = m {\delta v} \Delta({\delta v}).
\label{E_DeltaE}
\end{equation}

Solving (\ref{E_energy_balance})
for $\dot {\delta v}$ and replacing $\dot {\bar v} - g$ using (\ref{E_v}) gives 
\begin{equation}
\dot {\delta v} = ({\bar v}/{\delta v}) \dot N_{\rm w} \Delta {\bar v}/N
- \dot N_{\rm g} \Delta({\delta v})/N .
\label{E_sigma}
\end{equation}
As for the average velocity, (\ref{E_v}), the typical relative
velocity has a gain and a loss term. The gain term has a remarkable
symmetry to the loss term in (\ref{E_v}), which is completely
general. Only the second term in (\ref{E_sigma}) may be different, if
additional modes of energy dissipation like collisions with the walls
or friction are included.
The gain term in (\ref{E_sigma}) subsumes also the production of
granular temperature in the interior of the pipe due to the finite shear
rate, which is remarkable, as it expresses everything in terms of
physics at the wall.

Once the loss terms of the balance equations, (\ref{E_v}) and
(\ref{E_sigma}), are known, the time evolution of the average velocity
and the velocity fluctuations can be calculated, because the gain terms
are given. In this sense, it is sufficient to have a statistical
description of collisional cooling (which gives the loss term in
(\ref{E_sigma})) and of the momentum transfer of the granular gas to
a wall (which gives the loss term in (\ref{E_v})) in 
order to describe granular flow in a vertical pipe.
It turns out, that collisional cooling is easier, because it cannot
depend on the average velocity due to the
Galilei invariance of the grain-grain-interactions, whereas 
the momentum transfer to the walls depends on both, ${\bar v}$ and
$\delta v$.

\section{Collisional cooling}

We shall now specify $\dot N_{\rm g}$ and $\Delta({\delta v})$. 
The time between
two subsequent collisions of a particle can be estimated by the mean
free path, $\lambda$, divided by a typical relative velocity, 
${\delta v}$. Hence  
the number of binary collisions per unit time is proportional to
\begin{equation}
\dot N_{\rm g} \propto N {\delta v}/ \lambda .
\label{E_coll.rate0}
\end{equation}
Here we assumed that the flow is sufficiently homogeneous, that the
local variations of $\lambda$ and ${\delta v}$  are unimportant. 

In each collision the relative normal velocity gets
reduced by a factor, the restitution coefficient
$e_{\rm n}<1$. For simplicity we assume that the restitution
coefficient is a constant. Correspondingly a fraction of the
kinetic energy of relative motion is dissipated in each collision.
\begin{equation}
\Delta E = (1-e_{\rm n}^2){m\over2}{\delta v}^2
\label{E_restitution}
\end{equation}
with the grain mass $m$. According to (\ref{E_DeltaE}) $\Delta({\delta
  v}) = (1-e_{\rm n}^2) {\delta v}/2$. Putting this together, the
  dissipation rate (\ref{E_E_diss}) is \cite{Haff}
\begin{equation}
\dot E_{\rm diss} = k_{\rm g} N {m\over d} {\delta v}^3.
\label{E_E_diss1}
\end{equation}
The dimensionless proportionality constant $k_{\rm g}$ contains the
dependence on the solid fraction $\nu \propto d/\lambda$ and the
restitution coefficient $e_{\rm n}$ and
can be calculated analytically, if one assumes that the probability
distribution of the particles is Gaussian \cite{Lun}.

From these considerations one can draw a very general conclusion for
the steady state values of ${\bar v}$ and
${\delta v}$. In the steady state the kinetic 
energy is constant, so that (\ref{E_energy_balance}) together with
(\ref{E_E_pot}) and (\ref{E_E_diss1}) implies
\begin{equation}
{{{\bar v}_{\rm s}}\over {{\delta v}_{\rm s}^3}} = {{k_{\rm g}}\over {gd}}
\label{scaling1}
\end{equation}
Whenever the dissipation is dominated by irreversible binary collisions
and the flow is sufficiently homogeneous,
the steady flow velocity in a vertical pipe should be proportional to 
the velocity fluctuation to the power $3/2$. The proportionality 
constant does not depend on the width of the pipe. 

We tested this relation by two dimensional event driven molecular
dynamics simulations \cite{Wolf}. The agreement is surprisingly
good, given the simple arguments above, even quantitatively.
However, it turns out that the
proportionality constant in (\ref{scaling1}) has a weak
dependence on the width of the pipe, which can be traced back to
deviations of the velocity distribution from an isotropic Gaussian:
The vertical velocity component has a skewed distribution with
enhanced tail towards zero velocity \cite{Schaefer}.

\section{Interaction of the granular flow with the wall}

The collision rate $\dot N_{\rm w}$ with the walls of the vertical pipe
can be determined by noting that the number of particles colliding with
a unit area of the wall per unit time for low density $n$ is given by
$|v_{\perp}| n$. As the typical velocity perpendicular to the pipe
wall, $|v_{\perp}|$ is proportional to $\delta v$, one obtains
\begin{equation}
\dot N_{\rm w} \propto N {\delta v}/W.
\end{equation}
This is the place where the pipe width $W$ enters into the flow
dynamics. 

To specify, by how much the vertical velocity of a grain changes on
average, when it collides with the wall, is much more difficult, as it
depends on the local geometry of the wall. In our simulations the 
wall consisted of a dense array of circular particles of equal size.
When a grain is reflected from such a wall particle, a fraction of the
vertical component of its velocity will be reversed.
Instead of averaging this over all collision geometries, we
give some general arguments narrowing down the possible functional form
of $\Delta {\bar v}$. If we assume that the velocity distribution is
Gaussian, all moments of any velocity component must be functions of
${\bar v}$ and ${\delta v}$. This must be true for $\Delta {\bar v}$,
as well. For dimensional reasons it must be of the form 
\begin{equation}
\Delta {\bar v} = {\bar v} f\left({{\delta v}\over {\bar v}}\right)
\label{E_DeltaV}
\end{equation}
with a dimensionless function $f$. The physical interpretation of this
is the following: The loss term in the momentum balance can be viewed as
an effective wall friction. As long as the granular 
flow in the vertical pipe approaches a steady state, the friction force
must depend on the velocity ${\bar v}$. The ratio ${\delta v}/{\bar
  v}$ can be viewed as 
a characteristic impact angle, so that the function $f$ contains the
information about the average local collision geometry at the wall.
In principle all dimensionless parameters
characterizing the system may enter the funcion $f$, that is, apart
from $\nu$ also the
restitution coefficient $e_{\rm n}$ and the ratios $W/d$ and $gd/{\bar
  v}^2$. However, it is hard 
to imagine, that the width $W$ of the pipe or the gravitational
acceleration $g$ influences the local collision
geometry. Therefore we shall assume that $f$ does not depend
on $W/d$ or $gd/{\bar v}^2$. On the other hand, it is plausible, that 
the restitution coefficient enters $f$. It will influence the spatial 
distribution of particles and also accounts for the correlation 
of the velocities, if some particle is scattered back and forth 
between the wall and neighboring particles inside the pipe, and hence
hits the wall twice or more times without a real randomization of its
velocity. Due to positional correlations among the particles, $f$ should also
depend on the solid fraction $\nu$: One can easily imagine, that the
average collision geometry is different in dense and in
dilute systems.

Lacking a more precise understanding of the function $f$ we make a 
simple power law ansatz for it and write the loss term of
(\ref{E_v}) as
\begin{eqnarray}
{{\dot N_{\rm w}}\over N} \Delta {\bar v}  &=& 
{1\over W}\, {\delta v} {\bar v} \, k_{\rm w}\left({{\delta v}\over
{\bar v}}\right)^{\beta}\nonumber \\
 &=& k_{\rm w}\, W^{-1} {\delta v}^{1+\beta} {\bar v}^{1-\beta}.
\end{eqnarray}
The dimensionless parameters $k_{\rm w}$ and $\beta$ will be
functions of $\nu$ and $e_{\rm n}$. 

\section{Time evolution and steady state}

With these assumptions, the equations of motion (\ref{E_v}) and
(\ref{E_sigma}) for granular flow through a vertical pipe become
\begin{eqnarray}
\label{dot_v}
\dot {\bar v} &=& g -  k_{\rm w}\, W^{-1} {\delta v}^{1+\beta} {\bar v}^{1-\beta},\\
\label{dot_sigma}
\dot {\delta v} &=& k_{\rm w}\, W^{-1} {\delta v}^{\beta} {\bar v}^{2-\beta}
- k_{\rm g}\, d^{-1} {\delta v}^2.
\end{eqnarray}
As the time evolution should not be singular
for ${\bar v}=0$ or ${\delta v} = 0$, the values of $\beta$ are restricted to the
interval
\begin{equation}
0\leq \beta \leq 1 .
\label{E_Einschraenkung}
\end{equation}

The meaning of the exponent $\beta$ becomes clear, if we calculate
the steady state velocity from (\ref{dot_v}) and (\ref{scaling1}). The
result is 
\begin{equation}
{\bar v}_{\rm s} = \sqrt{gd}\, k_{\rm g}^{\gamma -1/2}  k_{\rm w}^{-\gamma}\,
\left({W\over d}\right)^{\gamma}.
\label{v_steady}
\end{equation} 
The exponent $\gamma$, which determines the dependence of the average
flow velocity on the pipe diameter, is related to $\beta$ by
\begin{equation}
\gamma = {3\over 2(2-\beta)}.
\end{equation}
Due to (\ref{E_Einschraenkung}) we predict that in granular pipe flow
\begin{equation}
3/4 \leq \gamma \leq 3/2,
\end{equation}
as long as the flow is sufficiently homogeneous and the main dissipation
mechanism are binary collisions. Note, that the exponent is always
smaller than 2, which would be its value for Hagen-Poisseuille flow
of a Newtonian fluid. $\gamma=3/2$ is the prediction of Bagnold's
theory, but in our simulations we found also values as small as 1,
depending on the values of the solid fraction and the restitution
coefficient \cite{Schaefer}. 

The stationary value ${\delta v}_{\rm s}$ directly follows from
(\ref{v_steady}) and (\ref{scaling1}). One obtains the same formula as
(\ref{v_steady}) with $\gamma$ replaced by $\gamma/3$.

\section{Collisional cooling for monopolar charged grains}

In this section we summarize our recent results
\cite{Scheffler1,Scheffler2}, how the dissipation 
rate (\ref{E_E_diss1}) is changed if all grains carry
the same electrical charge $q$ (besides having the same mass $m$,
radius $r$ and restitution coefficient $e_{\rm n}$). For simplicity we
assume that the charges are located in the middle of the insulating
particles. The results are valid for grains in a three
dimensional space, $D=3$. 

Whereas the hard sphere gas has no characteristic energy scale,
the Coulomb repulsion introduces such a scale,
\begin{equation}
E_{\rm q} = q^2/d.
\end{equation}
It is the energy barrier that two collision partners have to overcome,
when approaching each other from infinity. It has to be
compared to the typical kinetic energy stored in the relative motion
of the particles, which by 
analogy with molecular gases is usually expressed in terms of the
``granular temperature''
\begin{equation}
T = \delta v^2/D. 
\label{eq:T}
\end{equation}
If $E_{\rm q} \ll m\,T$ one expects that the charges have negligible
effect on the dissipation rate.

Using (\ref{eq:T}) and the expression 
\begin{equation}
\nu = \frac{\pi}{6} n d^3  \quad {\rm with}\quad n=N/V
\end{equation}
for the three dimensional solid fraction, the dissipation rate (\ref{E_E_diss1})
can be written in the form
\begin{equation}
\dot E_{\rm diss}/V = k \, n^2 d^2 m T^{3/2}
\label{eq:E_diss2}
\end{equation}
with the dimensionless prefactor
\begin{equation}
k = k_{\rm g} \pi \sqrt{3}/2\nu.
\label{eq:k}
\end{equation}
The advantage of writing it this way is that the leading $n$- or
$\nu$-dependence is explicitely given:
In the dilute limit $\nu \rightarrow 0$ the dissipation rate should be
proportional to $n^2$, i.e. to the probability that two particles meet in an
ideal gas. 

Since the remaining factors in (\ref{eq:E_diss2}) are uniquely determined 
by the dimension of the dissipation rate, this equation must hold for
charged particles as well. However, in this case the prefactor $k$
will not only depend on $e_{\rm n}$ and $\nu$, but also on the
dimensionless energy ratio $E_{\rm q}/mT$. We found
\cite{Scheffler1,Scheffler2} that the following factorization holds
\begin{equation}
k = k_0 (e_{\rm n}) g_{\rm chs}(\nu, E_{\rm q}/m T),
\end{equation}
where
\begin{equation}
k_0 = 2\sqrt{\pi}(1-e_{\rm n}^2)
\label{eq:k0}
\end{equation}
is the value of $k$ for $\nu=E_{\rm q}/mT=0$.
$g_{\rm chs}$ denotes the radial distribution function for charged
hard spheres (chs) at contact, normalized by the one for the uncharged ideal gas.

For $\nu < 0.2$ and  $E_{\rm q}/mT < 8$ our computer simulations show that
\begin{equation}
g_{\rm chs}\left(\nu,\frac{E_{\rm q}}{m T}\right) \approx g_{\rm
  hs}(\nu) \exp\left(-\frac{E_{\rm q}}{m T}f(\nu)\right)  
\label{eq:g}
\end{equation}
is a very good approximation. Here, 
\begin{equation}
g_{\rm hs} = \frac{2-\nu}{2(1-\nu)^3} \geq 1
\end{equation}
is the well-known Enskog correction for the radial distribution
function of (uncharged) hard spheres (hs) \cite{CarStar}.
This factor describes that the probability that two particles collide
is enhanced due to the excluded volume of all the remaining particles.
The second, Boltzmann-like factor describes that the Coulomb repulsion
suppresses collisions. The effective energy barrier $E_{\rm q} f(\nu)$
decreases with increasing solid fraction, because two particles which
are about to collide not only repel each other but are also pushed
together by being repelled from all the other charged particles in the
system. A two parameter fit gives
\begin{equation}
f(\nu) \approx  1 - c_0 \, \nu^{1/3} + c_1 \, \nu^{2/3} 
\end{equation}
with
\begin{equation}
c_0 = 2.40 \pm 0.15, \quad {\rm and} \quad c_1 = 1.44 \pm 0.15 .
\label{eq:fit}
\end{equation}
Very general arguments \cite{Scheffler1} lead to the prediction that
$c_1 = (c_0/2)^2$, which is confirmed by (\ref{eq:fit}).

We expect deviations from (\ref{eq:g}) for larger $\nu$ and $E_{\rm
  q}/mT$, because the uncharged hard sphere system has a fluid-solid
  transition close to $\nu \approx 0.5$, and the charged system
may become a Wigner crystal for any solid fraction, provided the
  temperature gets low enough.

\section{Conclusion}
 
We presented four main results: The steady state velocity of granular
flow in a vertical pipe should have
a power law dependence on the diameter $W$ of the pipe with an exponent $\gamma$
ranging between 3/4 and 3/2, depending on the solid fraction and the 
restitution coefficient of the grains. This result was derived
ignoring possible electric charges of the grains and assuming that the
flow is sufficiently homogeneous and the main dissipation mechanism
are binary collisions. This illustrates the genuinely non-Newtonian
character of granular flow. 

Second, the dependence of the steady state velocity on the
solid fraction $\nu$, the restitution coefficient $e_{\rm n}$ and --
in the case of monopolar charging -- the ratio between Coulomb barrier
and kinetic energy,  $E_{\rm q}/mT$, is contained in the factor
$k_{\rm g}^{\gamma -1/2} k_{\rm w}^{-\gamma}$ in (\ref{v_steady}).
In the dilute limit $\nu \rightarrow 0$ as well as in the limit of
nearly elastic particles $e_{\rm n} \rightarrow 1$ the coefficient
$k_{\rm w}$, which describes how sensitive the momentum transfer
to the wall depends on the local collision geometry, should remain
finite, whereas $k_{\rm g}$ vanishes like $\nu (1-e_{\rm n}^2)$
according to  (\ref{eq:k}), (\ref{eq:k0}). As $\gamma-1/2 > 0$, this
implies that the flow through a vertical pipe becomes faster the higher
the solid fraction and the less elastic the collisions between the
grains are (in the limit of low density and nearly elastic
collisions). The physical reason for this is that in denser and more
dissipative systems collisional cooling is more efficient,
reducing the collisions with the walls and hence their braking effect.
This remarkable behaviour has been confirmed in computer simulations
\cite{Schaefer}.

Third, monopolar charging leads to a Boltzmann-like factor in $k_{\rm
  g}$ or the dissipation rate, respectively, which means that for low
granular temperature the dissipation rate becomes exponentially weak.
The higher the density the less pronounced is this effect, because the
effective Coulomb barrier $E_{\rm q} f(\nu)$ hindering the collisions
  becomes weaker. 

Finally, we derived the evolution equations for the flow velocity and
the velocity fluctuation for granular flow through a vertical pipe,
(\ref{dot_v}) and (\ref{dot_sigma}). These equations apply to the
situation of homogeneous flow, which can only be realized in computer
simulations of a sufficiently short pipe with periodic boundary
conditions. In order to generalize these equations for flow that is
inhomogeneous along the pipe one should replace the time derivatives
by $\partial_t + \bar{v}(z,t) \partial_z$. In addition a third equation,
the continuity equation, is needed to describe the time evolution of
the density of grains along the pipe. Such equations have been
proposed previously \cite{Lee,Valance,Riethmuller} in order to study the kinetic 
waves spontaneously forming in granular pipe flow. Our equations are
different.

\section*{Acknowledgements}
We thank the Deutsche Forschungsgemeinschaft for supporting this
research through grant no. Wo 577/1-3. The computer simulations
supporting the theory presented in this paper were partly done at the
John von Neumann-Institut f\"ur Computing (NIC) in J\"ulich.

\end{document}